\documentstyle[12pt]{article}

 \font\tenmsb=msbm10 scaled\magstep 1
   \font\sevenmsb=msbm7 scaled \magstep 1
   \font\faivemsb=msbm5 scaled \magstep 1
\newfam\msbfam
      \textfont\msbfam=\tenmsb
      \scriptfont\msbfam=\sevenmsb
      \scriptscriptfont\msbfam=\faivemsb
\def\Bbb#1{{\fam\msbfam #1}}
\font\tengothic=eufm10 scaled\magstep 1
\font\sevengothic=eufm7 scaled\magstep 1
\newfam\gothicfam
      \textfont\gothicfam=\tengothic
      \scriptfont\gothicfam=\sevengothic

\newcommand{\lbd}{\lambda}
\newcommand{\Lbd}{\Lambda}
\newcommand{\be}{\begin{equation}}
\newcommand{\ee}{\end{equation}}
\newcommand{\dlt}{\delta}
\newcommand{\Dlt}{\Delta}
\newcommand{\Gm}{\Gamma}
\newcommand{\gm}{\gamma}
\newcommand{\sgm}{\sigma}
\newcommand{\ra}{\rightarrow}
\newcommand{\al}{\alpha}
\newcommand{\om}{\omega}
\newcommand{\Om}{\Omega}
\newcommand{\prt}{\partial}

\begin{document}

\begin{center}
{\large{\bf Optical Turbulent Structures} \\ [5mm]
V.I. Yukalov}

{\it Bogolubov Laboratory of Theoretical Physics \\
Joint Institute for Nuclear Research, Dubna 141980, Russia}

\end{center}

\vskip 1cm

\begin{abstract}

The problem of describing optical turbulent structures, arising in resonant 
media with high Fresnel numbers, is reviewed. The consideration is based on 
the probabilistic approach to pattern selection, ascribing a probability 
distribution of patterns to systems with multiple possible structures. The 
most probable structure corresponds to the minimal expansion exponent, or to 
the minimal expansion rate, characterizing the phase-space expansion of a 
dynamical system. Turbulent photon filamentation is studied. The most probable 
filament radius and the number of filaments are found, being in good agreement 
with experiment.

\end{abstract}

\vskip 1cm


\vskip 2.5cm

{\large {\bf 1. Experiments on Optical Structures}}

\vskip 3mm

\begin{sloppypar}

Spatio-temporal optical structures appear when electromagnetic
waves propagate through a medium possessing some sort of 
nonlinearity. One can distinguish three types of such media, 
depending on the physical origin of nonlinearity.

\end{sloppypar}

\vskip 1mm

(i) {\it Kerr medium}. This is a passive medium with a nonlinear 
relation between polarization and electric field, when susceptibility 
depends on the field intensity. Many liquid crystals pertain to this 
kind of matter. Different patterns arising in the Kerr medium are 
reviewed in Refs. [1--3].

\vskip 1mm

(ii) {\it Photorefractive medium}. This pertains to a medium with a
strong photorefractive effect. The photorefractive effect is the change 
in refractive index of a medium resulting from the optically induced 
redistribution of electrons and holes. The modulation of the medium 
refractive index is usually realized as a four-wave mixing process. 
Two input optical fields produce intensity grating, shifting electrons
from donor atoms, which induces a space charge wave, equivalent to the
formation of a ripple field. The nonlinear interaction of the input 
fields with the medium leads to the generation of two output waves.
An example of a photorefractive medium is the bismuth selicon oxide 
crystal Bi$_{12}$SiO$_{20}$. Various patterns are reviewed in Ref. [2,3].

\vskip 1mm

(iii) {\it Resonant medium}. This, evidently, is an active medium typical 
of laser systems. Nonlinearity arises because of an effective interaction
of resonant atoms through the common radiation field. Different 
spatio-temporal structures were observed in a Na$_2$ laser, CO$_2$ laser,
dye laser and in many semiconductor and vapour lasers, as is reviewed
in Refs. [3--6].

\vskip 1mm

Pattern formation in nonlinear optics, which is also called optical 
morphogenesis, has many similarities for different media. This especially
concerns the case when nonlinear optical samples have the same geometry.
The most often used is the cylindrical geometry typical of lasers.
Characteristic structures of cylindrical samples are filaments, aligned 
along the cylindrical axis, where the radiation intensity is much higher 
than in the space between these filaments. The appearance of such structures 
is easily noticeable in the transverse cross-section, where the high-intensity
filaments are exhibited as bright spots. For brevity, we call these 
filamentary structures {\it photon filaments} [5,6].

In the case of cylindrical samples, the peculiar properties of photon
filaments essentially depend on the Fresnel number
$$
F\equiv \frac{\pi R^2}{\lbd L} \; ,
$$
in which $R$ is the cylinder radius, $L$ is the characteristic length, 
and $\lbd$ is the radiation wavelength. The Fresnel number is a sort of an 
aspect ratio comparing the transverse and longitudinal characteristics.

The Fresnel number in optics plays the same role as the Reynolds number in
hydrodynamics. When increasing the Reynolds number, the fluid passes from 
laminar motion to turbulent one. A similar transition happens in optics when 
increasing the Fresnel number, and occurs around $F\approx 10$.

At small Fresnel numbers $F<5$, there can exist several transverse modes
with a regular behaviour. These transverse structures are regular in
space, forming ordered geometric arrays seen as polygons in the transverse
cross-section, with the number of bright spots proportional to $F^2$. The
structures are also regular in time being either stationary or periodically
oscillating. The type of such regular structures is prescribed by the sample
geometry and corresponds to the empty-cavity Gauss-Laguerre modes imposed by 
the cylindrical geometry. This regular behaviour in optics is analogous to 
the laminar motion in hydrodynamics, and it is well understood theoretically
both for nonlinear media, such as a Kerr medium and photorefractive crystals
[1--3], and for laser media [4].

For large Fresnel numbers $F>15$, the appearing spatio-temporal patterns 
are principally different from the empty-cavity modes. The modal expansion
is no longer valid and the boundary conditions have no importance. The 
medium looks as a bunch of independently flashing filaments, whose number 
is proportional to $F$. The filaments are chaotically distributed in the 
transverse cross-section, are not correlated with each other, and are 
flashing aperiodically in time. This spatio-temporal optical chaos is
similar to hydrodynamic turbulence, because of which it is called {\it
optical turbulence}. Since this phenomenon is characterized by the formation
of bright filaments with a high photon density, it has been named [7]
{\it turbulent photon filamentation}. This kind of optical turbulence was 
observed in nonlinear media, such as photorefractive crystals [2,3] as well 
as in different lasers, such as CO$_2$, dye, and vapour lasers (see discussion
in Refs. [5--7]).

The theory of turbulent photon filamentation was advanced in Ref. [7]. This 
theory is based on the probabilistic approach to pattern selection [6].

\vskip 5mm

{\large {\bf 2. Probabilistic Pattern Selection}}

\vskip 3mm

The problem of pattern selection arises when evolution equations possess 
several solutions corresponding to different spatio-temporal structures. This 
problem is rather common for many nonlinear systems, optical media being only
a particular case [8]. Therefore, let us, first, formulate a general approach
to pattern selection, relevant for any dynamical system.

Let us consider a set of functions, in general complex, $y_k(x,t)$ enumerated 
by the index $k=1,2,\ldots$ and depending on a collection of variables
$x\in\Bbb{D}\subset\Bbb{R}^d$ and on time $t\in\Bbb{R}$. The set of 
$y(t)\equiv\{ y_k(x,t)\}$ is the {\it dynamical state}. For the compactness of
notation, it is convenient to denote by one index $i$ the pair $k$ and $x$,
writing $y(t)=\{ y_i(t)\}$, with $y_i(t)\equiv y_k(x,t)$. Then the dynamical 
state $y(t)$ can be treated as a column with respect to $i$. In this notation, 
the system of evolution equations can be presented as
\be
\frac{d}{dt}\; y(t) = v(y,t) \; ,
\ee
with a velocity filed being also a column $v(y,t)=\{ v_i(y,t)\}$. Evolution
equations containing only the first derivatives in time are called to have
the {\it normal form}. Any system of equations containing higher time derivatives
can always be reduced to the normal form [9].

The dynamical state $y(t)\subset{\cal F}$ pertains to a {\it phase space}
${\cal F}=\otimes_i{\cal F}_i$, being a tensor product of the spaces 
${\cal F}_i\supset y_i(t)$. An {\it elementary phase volume} of the phase 
space ${\cal F}$ can be written [6,7] as $|\dlt\Gm(t)|$, with
\be
\dlt\Gm(t) \equiv \prod_i \dlt y_i(t) \; .
\ee
The temporal behaviour of the phase volume element (2) is characterized by the
{\it expansion exponent}
\be
\sgm(t) \equiv \ln\left | \frac{\dlt\Gm(t)}{\dlt\Gm(0)}\right | \; ,
\ee
which shows whether the phase volume expands with time or contracts according
to the law
$$
|\dlt\Gm(t)| = |\dlt\Gm(0)|e^{\sgm(t)} \; .
$$

The expansion exponent (3) can be presented in several equivalent forms [6,7].
Introducing the {\it multiplier matrix} $\hat M(t)=[M_{ij}(t)]$ with the elements
\be
M_{ij}(t) \equiv \frac{\dlt y_i(t)}{\dlt y_j(0)} \; ,
\ee
we can transform Eq. (3) to
\be
\sgm(t) = \ln|{\rm det}\; \hat M(t)| \; .
\ee

Another form of the expansion exponent (3) can be obtained by means of the {\it
expansion matrix} $\hat X(t)=[X_{ij}(t)]$ with the elements
\be
X_{ij}(t) \equiv \frac{\dlt v_i(y,t)}{\dlt y_j(t)} \; .
\ee
The variation of the evolution equation (1) yields
\be
\frac{d}{dt}\; \hat M(t) =\hat X(t)\hat M(t) \; .
\ee
>From Eqs. (6) and (7), we find
\be
\sgm(t) ={\rm Re}\int_0^t {\rm Tr}\hat X(t')\; dt' \; .
\ee

If the multiplier matrix (4) possesses eigenvalues $\mu_n(t)$ enumerated by a
multi-index $n$, then Eq. (5) gives
\be
\sgm(t) =\sum_n \ln|\mu_n(t)| \; .
\ee

Let us introduce the {\it local expansion rate}
\be
\Lbd(t) \equiv \frac{1}{t}\; \sgm (t) \; .
\ee
The latter, depending on the form of the expansion exponent, can be written in 
the following three ways:
\be
\Lbd(t) =\frac{1}{t}\; \ln\left |{\rm det}\hat M(t)\right | \; ,
\ee
\be
\Lbd(t) =\frac{1}{t}\; {\rm Re} \int_0^t {\rm Tr}\hat X(t') \; dt' \; ,
\ee
\be
\Lbd(t) = \sum_n \frac{1}{t}\; \ln\left | \mu_n(t) \right | \; .
\ee
>From Eq. (13), we have
\be
\lim_{t\ra\infty} \Lbd(t) = \sum_n \lbd_n \; ,
\ee
where the right-hand side is the sum of the Lyapunov exponents
\be
\lbd_n \equiv \lim_{t\ra\infty} \frac{1}{t}\; \ln\left | \mu_n(t) \right |\; ,
\ee
provided the corresponding limit exists.

It is easy to check that
\be
\lim_{t\ra\infty} \frac{d}{dt}\; \sgm(t) = \sum_n \lbd_n \; ,
\ee
which immediately follows from Eqs. (9) and (15). Since from Eq. (8) one has
$$
\frac{d}{dt}\; \sgm(t) = {\rm Re} \; {\rm Tr} \hat X(t) \; ,
$$
then, according to Eq. (16), one gets
$$
\lim_{t\ra\infty} {\rm Re} \; {\rm Tr} \hat X(t) = \sum_n \lbd_n \; .
$$

These limiting properties remind us those of the Gibbs entropy
$$
S_G(t) = -\int \rho(y,t)\ln\rho(y,t)\; dy \; ,
$$
in which $\rho(y,t)$ is a probability density. For the Gibbs entropy in steady
states, one has [10--14] the limit
$$
\lim_{t\ra\infty}\frac{d}{dt} \; S_G(t) = \sum_n \lbd_n \; .
$$
The quantity $dS_G/dt$ is called the entropy production rate or, simply, the
entropy production. Therefore, the expansion exponent $\sgm(t)$ is a kind of
{\it dynamic entropy variation}, and the expansion rate $\Lbd(t)$ resembles
the entropy production rate. More details on the entropy production can be
found in Refs. [11--16].

Notice that normally an invariant probability measure of a steady state is
singular [13] and therefore the Gibbs entropy $S_G(t)$ is $S_G(t)=-\infty$,
which is physically unacceptable.

It is worth mentioning that in dynamical theory the quantity 
${\it Re}\;{\rm Tr}\hat X(t)$ is sometimes termed the contraction rate.
According to Eq. (12), the expansion rate (10) could also be named the
contraction rate. The usage of the words "expansion" or "contraction" is
rather philological. But, probably, it is more logical to call $\Lbd(t)$ 
the expansion rate, since $\Lbd(t)>0$ means really the expansion of the 
phase volume $|\dlt\Gm(t)|$, while $\Lbd(t)<0$ implies the contraction of 
the latter.

Now let us turn to the problem of pattern selection, when the evolution
equation (1) possesses a set of solutions corresponding to different 
spatio-temporal structures. If we label these different patterns by an index
$\al$, then for each pattern we have a solution $y(\al,t)$ and, respectively,
an expansion exponent $\sgm(\al,t)$ and an expansion rate $\Lbd(\al,t)$.
The problem is how to classify these different solutions and the related 
patterns.

The probabilistic approach to pattern selection [6,7] is based on the idea 
that each pattern can be characterized by its probabilistic weight. That is,
there should exist a pattern distribution $p(\al,t)$. The latter can be 
derived by minimizing the {\it pattern information}
\be
I_p(t) \equiv \int p(\al,t) \ln p(\al,t)\; d\al +
\int p(\al,t)\sgm(\al,t) d\al \; ,
\ee
consisting of the sum of the Shanon information and lost information, 
under the normalization condition
$$
\int p(\al,t) \; d \al = 1 \; .
$$
This implies the minimization of the conditional pattern information
$$
\tilde I_p (t) \equiv I_p(t) +  l(t)\left [
\int p(\al,t)\; d\al - 1 \right ] \; ,
$$
in which $l(t)\equiv\ln Z(t) -1$ is a Lagrange multiplier. Minimizing 
$\tilde I_p(t)$ results in the {\it pattern distribution}
\be
p(\al,t) = \frac{1}{Z(t)} \; \exp\{ -\sgm(\al,t)\} \; ,
\ee
with the normalization factor
$$
Z(t) = \int \exp \{ -\sgm(\al,t)\} \; d\al \; .
$$
Taking into account the relation (10), we obtain
\be
p(\al,t) =\frac{1}{Z(t)} \; \exp\left\{ - \Lbd(\al,t)\; t \right \} \; .
\ee
In this way, the {\it most probable pattern} corresponds to the
{\it minimal expansion exponent} $\sgm(\al,t)$ or, equivalently, to the
{\it minimal expansion rate} $\Lbd(\al,t)$, that is, one has the relation
\be
\min_\al \Lbd(\al,t) \Longleftrightarrow \max_\al p(\al,t) \; .
\ee
Thus, we come to the following conclusion.

\vskip 2mm

{\bf Principle of Pattern Selection}. The most probable pattern for a
dynamical system at a given time is defined by the minimal expansion rate.

Let us emphasize that the formulated here approach of {\it Probabilistic
Pattern Selection} is a very different from the Prigogine [17--21] principle
of minimal entropy production in stationary states. The latter principle 
deals with the entropy production $\dot{S}\equiv dS/dt$ defined as time 
derivative of thermodynamical entropy. It says that the entropy production
$\dot{S}(\al)$ in a stationary state is minimal with respect to internal
thermodynamic variables $\al$. The basic points, distinguishing the
{\it Probabilistic Pattern Selection} from the Prigogine principle, are as 
follows:

\vskip 1mm

(i) This approach is developed for arbitrary dynamical systems. In 
particular, it may concern the evolution equations for thermodynamic 
variables or for statistical averages of some operators.

\vskip 1mm

(ii) The theory is valid for any dynamical states, not necessarily stationary
ones.

\vskip 1mm

(iii) Not solely the most probable pattern is defined, but the 
probability distribution for all feasible patterns is suggested.

\vskip 5mm

{\large {\bf 3. Turbulent Photon Filamentation}}

\vskip 3mm

Now we shall apply the theory of Section 2 to the problem of turbulent 
photon filamentation, discussed in Section 1. Consider a system of $N$ 
resonant atoms interacting with electromagnetic field as is described
by the Hamiltonian
\be
\hat H = \hat H_a + \hat H_f + \hat H_{af}
\ee
consisting of the terms: The Hamiltonian of two-level resonant atoms
\be
\hat H_a = \sum_{i=1}^N \om_0 \left ( \frac{1}{2} +
S_i^z \right ) \; ,
\ee
where $\om_0$ is a transition frequency and $S_i^z$ is a spin-$1/2$
operator. The radiation-field Hamiltonian
\be
\hat H_f = \frac{1}{8\pi} \int \left ( {\bf E}^2 + {\bf H}^2
\right ) \; d{\bf r} \; ,
\ee
with electric field ${\bf E}$ and magnetic field ${\bf H}=
{\bf\nabla}\times{\bf A}$. The vector potential is assumed to satisfy the
Coulomb gauge calibration ${\bf\nabla}\cdot{\bf A}=0$. The atom-field 
interaction Hamiltonian
\be
\hat H_{af} = - \sum_{i=1}^N \left (
\frac{1}{c} {\bf J}_i\cdot {\bf A}_i + {\bf D}_i \cdot {\bf E}_{0i}
\right ) \; ,
\ee
in which ${\bf A}_i \equiv {\bf A}({\bf r},t)$, and the transition-current
and transition-dipole operators are 
$$
{\bf J}_i = i\om_0 \left ( {\bf d} S_i^+ - {\bf d}^* S_i^- \right ) \; ,
\qquad {\bf D}_i = {\bf d} S_i^+ + {\bf d}^* S_i^- \; ,
$$
where ${\bf d}$ is the transition dipole, ${\bf d}\equiv d_0{\bf e}_d$;
$S_i^\pm$ are the rising or lowering operators, respectively; and 
${\bf E}_{0i}$ is a seed field.

We shall consider the evolution equations for the functions
\be
u({\bf r},t) \equiv 2 < S^-({\bf r},t)>\; , \qquad
s({\bf r},t) \equiv 2 < S^z({\bf r},t)>\; ,
\ee
being the statistical averages of the corresponding operators. Introduce the
notation
\be
f({\bf r},t) = f_0({\bf r},t) + f_{rad}({\bf r},t)
\ee
for an effective field acting on an atom, with the cavity seed field
\be
f_0({\bf r},t) \equiv - 2i{\bf d}\cdot {\bf E}_0({\bf r},t)
\ee
and the radiation field
\be
f_{rad}({\bf r},t) \equiv -\; \frac{3}{4}\; i\gm\rho 
\int \left [ \frac{e^{ik_0|{\bf r}-{\bf r}'|}}{k_0|{\bf r}-{\bf r}'|}\;
u({\bf r}',t) - {\bf e}_d^2\; 
\frac{e^{-ik_0|{\bf r}-{\bf r}'|}}{k_0|{\bf r}-{\bf r}'|}\; u^*({\bf r}',t)
\right ]\; d{\bf r}' \; ,
\ee
where $k_0\equiv\om_0/c$, $\rho$ is the density of atoms, and 
$\gm\equiv 4k_0^3d_0^2/3$ is the natural width. The seed field
\be
{\bf E}_0 = \frac{1}{2}\; {\bf E}_1\; e^{i(kz-\om t)} +
\frac{1}{2}\; {\bf E}_1^*\; e^{- i(kz-\om t)}
\ee
selects the longitudinal mode with the frequency $\om=kc$ satisfying the
quasiresonance condition
$$
\frac{|\Dlt|}{\om_0} \ll 1 \; , \qquad \Dlt \equiv \om -\om_0\; .
$$

The evolution equations for the atomic variables (25) can be derived [5]
by eliminating the field variables and invoking the semiclassical 
approximation. This yields 
$$
\frac{\prt u}{\prt t} = - (i\om_0 +\gm_2) u + f s\; , \qquad
\frac{\prt |u|^2}{\prt t} = - 2\gm_2 |u|^2 + (u^* f + f^* u) s \; ,
$$
\be
\frac{\prt s}{\prt t} = -\; \frac{1}{2} \left ( u^* f + f^* u\right ) -
\gm_1 ( s - \zeta ) \; ,
\ee
where $\gm_1$ is the level width, $\gm_2$ is the line width, and $\zeta >0$
is a pumping parameter.

Without the loss of generality, we may look for the solutions to Eqs. (30)
in the form of a bunch of $N_f$ filaments,
\be
u({\bf r},t) = \sum_{n=1}^{N_f} u_n(r_\perp,t) e^{ikz} \; , \qquad
s({\bf r},t) = \sum_{n=1}^{N_f} s_n(r_\perp,t) \; ,
\ee
with $r_\perp\equiv\sqrt{x^2+y^2}$. In the case of very small Fresnel
numbers, $F\ll 1$, there is the sole filament, so that $N_f=1$ and $u_n$ 
and $s_n$ are uniform in space. For low Fresnel numbers $F<10$, when the 
regime of laminar filamentation occurs, an expansion over Gauss-Laguerre
modes is more appropriate [4]. And for large Fresnel number $F>10$, when
the regime of turbulent filamentation develops, there appear many 
uncorrelated filaments. Then the functions $u_n$ and $s_n$ are essentially
nonzero only around the axis of an $n$-th filament, but fastly decrease
outside the filament, so that $u_mu_n\sim\dlt_{mn}$, $s_ms_n\sim\dlt_{mn}$,
and $u_ms_n\sim\dlt_{mn}$. Here we are interested in the latter case of 
turbulent photon filamentation.

To describe the turbulent regime, let us introduce the averaged functions
\be
u_n(t) \equiv \frac{1}{V_n}\; \int_{V_n} u_n(r_\perp,t) \; d{\bf r} \; ,
\qquad
s_n(t) \equiv \frac{1}{V_n}\; \int_{V_n} s_n(r_\perp,t) \; d{\bf r} \; ,
\ee
with the averaging over a cylinder enveloping the $n$-th filament. The
volume of the enveloping cylinder is $V_n=\pi a_n^2 L$, with $a_n$ being
the cylinder radius and $L$, the sample length. The radius of the enveloping
cylinder $a_n$ is related to the filament radius $r_n$ by the 
{\it conservation-energy equation}
\be
\int | u_n(r_\perp,t) |^2 \; d{\bf r} = V_n | u_n(r_n,t) |^2 \; .
\ee
If the filament profile is well approximated by the normal law 
$\exp(-r_\perp^2/2r_n^2)$, the relation (33) gives
\be
a_n = 1.82 \; r_n \; .
\ee

The sample is assumed to have the cylindrical shape of radius $R$ and length
$L$. For these values and for the radiation wavelength $\lbd$, the inequalities
$$
\lbd \ll R \ll L \; ,
$$
typical of lasers, are valid. For the functions (32), characterizing the
turbulent regime, we obtain the equations
$$
\frac{du_n}{dt} = -(i\Om_n +\Gm_n) u_n + f_1 s_n \; , \qquad
\frac{d|u_n|^2}{dt} =  -2\Gm_n|u_n|^2 + (u_n^* f_1 + f_1^* u_n) s_n \; ,
$$
\be
\frac{ds_n}{dt} = - g_n\gm_2 |u_n|^2 -\; \frac{1}{2} \left ( u_n^* f_1 +
f_1^* u_n\right ) - \gm_1 (s_n -\zeta) \; ,
\ee
where
$$
\Gm_n \equiv \gm_2(1-g_ns) \; , \qquad \Om_n\equiv \om_0 +\gm_2 g_n' \; ,
\qquad f_1 \equiv -i{\bf d}\cdot{\bf E}_1 e^{-i\om t} \; ,
$$
and the efective {\it coupling parameters} are
\be
g_n \equiv \frac{3\gm\rho}{4\gm_2} \int_{V_n} 
\frac{\sin(k_0r - kz)}{k_0 r}\; d{\bf r} \; , \qquad 
g_n' \equiv \frac{3\gm\rho}{4\gm_2} \int_{V_n} 
\frac{\cos(k_0r - kz)}{k_0 r}\; d{\bf r} \; .
\ee
The equations for different filaments are naturally decoupled since the 
filaments are not correlated with each other.

The nonlinear system of equations (35) can be solved by means of the scale
separation approach [5,22]. The details of this solution can be found in 
Refs. [6,7]. Calculating the expansion rate (12) and minimizing it with 
respect to the filament radius, we obtain the most probable radius
\be
r_f = 0.3\sqrt{\lbd L} \; ,
\ee
which is in good agreement with experiment, as is discussed in refs. [5--7]. 
The formula (37) has been specially checked in the regime of turbulent photon
filamentation of CO$_2$ and dye lasers and found to well describe the
experimental observations [23--27].

The most probable number of filaments can be estimated from the normalization
integral
\be
\frac{1}{V} \int s\; d{\bf r} = \zeta \; ,
\ee
where the integration is over the whole volume of the sample $V=\pi R^2L$. 
Considering the developed turbulent regime, when the population difference 
inside each filament of radius $r_f$ has reached the value of the pumping
parameter $\zeta$ and the population difference outside the filaments is
$s_{out}$, we have from the integral (38) the equation
$$
N_f V_f \zeta + (V - N_f V_f) s_{out} = \zeta V\; .
$$
>From here, we find
\be
N_f = \frac{V}{V_f} = \left ( \frac{R}{r_f}\right )^2 \; .
\ee
With the filament radius (37), this gives
\be
N_f\cong 4F \; , \qquad \frac{r_f}{R} \simeq \frac{1}{2\sqrt{F}}\; ,
\ee
which also is in good agreement with experiment.

Filamentation in laser media is a rather general process and should arise
for different samples, provided the Fresnel number is sufficiently high.
This should concern as well the so-called random lasers [28].

The theory presented above has to do with the well developed optical 
turbulence corresponding to large Fresnel numbers. An interesting question
is how to describe the intermediate region, where $5\leq F\leq 15$, and
the arising turbulence is yet weak, being intermittent with the remnants of 
the regular filaments, characterized by the Gauss-Laguerre modes. Then between 
the laminar and turbulent filaments there could appear a kind of Josephson 
interference, as it happens for electromagnetic vortices [29,30]. In the 
frame of the probabilistic pattern selection, described in Sec. 2, we could
treat laminar and turbulent filaments as independent, calculating their
probabilities. In the intermediate region of Fresnel numbers, where these
two types of filaments coexist, their probabilities should be comparable.
At low Fresnel numbers, the probability of the laminar filament must prevail,
while at high $F$, the probability of the turbulent filaments  is to become
much larger. In this way, the whole picture for arbitrary $F$, varying in
the interval $[0,\infty)$, can be done in the frame of the probabilistic
pattern selection.


\begin{thebibliography}{99}
\bibitem{1}
S.A. Akhmanov, R.V. Khokhlov, and A.P. Sukhorukov, in {\it Laser Handbook},
edited by F.T. Arecchi and E. Shultz-DuBois (North-Holland, Amsterdam, 1972),
V. 2, p. 1151.

\bibitem{2}
F.T. Arecchi, Physica D {\bf 86}, 297 (1995).

\bibitem{3}
F.T. Arecchi, S. Boccaletti, and P.L. Ramazza, Phys. Rep. {\bf 318}, 1 (1999).

\bibitem{4}
L.A. Lugiato, Phys. Rep. {\bf 219}, 293 (1992).

\bibitem{5}
V.I. Yukalov and E.P. Yukalova, Phys. Part. Nucl. {\bf 31}, 561 (2000).

\bibitem{6}
V.I. Yukalov, Physica A {\bf 291}, 255 (2001).

\bibitem{7}
V.I. Yukalov, Phys. Lett. A {\bf 278}, 30 (2000).

\bibitem{8}
M.C. Cross and P.C. Hohenberg, Rev. Mod. Phys. {\bf 65}, 851 (1993).

\bibitem{9}
N.N. Bogolubov and B.I. Sadovnikov, Mosc. Univ. Phys. Bull. {\bf 3}, 24 (1961).

\bibitem{10}
N.I. Chernov, G.I. Eyink, J.L. Lebowitz, and Y.G. Sinai, Commun. Math. Phys.
{\bf 154}, 569 (1993).

\bibitem{11}
D. Ruelle, J. Stat. Phys. {\bf 85}, 1 (1996).

\bibitem{12}
W. Breymann, T. Tel, and J. Volmer, Phys. Rev. Lett. {\bf 77}, 2945 (1996).

\bibitem{13}
D. Ruelle, J. Stat. Phys. {\bf 95}, 393 (1999).

\bibitem{14}
L. Rondoni and E.G.D. Cohen, Nonlinearity {\bf 13}, 1905 (2000).

\bibitem{15}
R. Luzzi and A.R. Vasconcellos, Fortschr. Phys. {\bf 38}, 887 (1990).

\bibitem{16}
J.G. Ramos, A.R. Vasconcellos, and R. Luzzi, Fortschr. Phys. {\bf 43}, 265 
(1995).

\bibitem{17}
I. Prigogine, {\it Introduction to Nonequilibrium Thermodynamics} (Wiley,
New York, 1962).

\bibitem{18}
I. Prigogine, {\it Nonequilibrium Statistical Mechanics} (Wiley,
New York, 1962).

\bibitem{19}
P. Glansdorff and I. Prigogine, {\it Thermodynamic Theory of Structure, 
Stability and Fluctuations} (Wiley, New York, 1971).

\bibitem{20}
W. Ebeling, {\it Structurbildung bei Irreversiblen Prozessen} (Teubner, 
Leipzig, 1976).

\bibitem{21}
Y.L. Klimontovich, {\it Turbulent Motion and Structure of Chaos} (Nauka, 
Moscow, 1990).

\bibitem{22}
V.I. Yukalov, Phys. Rev. B {\bf 53}, 9232 (1996).

\bibitem{23}
I. Pastor and J.M. Guerra, Appl. Phys. B {\bf 51}, 342 (1990).

\bibitem{24}
I. Pastor, V.M. P\'erez-Garcia, F. Encinas-Sanz, J.M. Guerra, and L. 
Vazquez, Physica D {\bf 66}, 412 (1993).

\bibitem{25}
V.M. P\'erez-Garcia and J.M. Guerra, Phys. Rev. A {\bf 50}, 1646 (1994).

\bibitem{26}
V.M. P\'erez-Garcia, I. Pastor, and J.M. Guerra, Phys. Rev. A {\bf 52}, 2392 
(1995).

\bibitem{27}
F. Encinas-Sanz, J.M. Guerra, and I. Pastor, Opt. Lett. {\bf 21}, 1153 (1996).

\bibitem{28}
X. Jiang and C.M. Soukoulis, Phys. Rev. Lett. {\bf 85}, 70 (2000).

\bibitem{29}
A. Vourdas, Europhys. Lett. {\bf 48}, 201 (1999).

\bibitem{30}
A. Vourdas, Phys. Rev. B {\bf 60}, 620 (1999).

\end{thebibliography}
\end{document}